# *A Data Aggregation Scheme Based on Secret Sharing with Error Tolerance in Smart Grid*

Zhitao Guan[1], Guanlin Si[1], Xiaojiang Du[2], Peng Liu[3]

1. School of Control and Computer Engineering, North China Electric Power University, China, guan@ncepu.edu.cn,
2. Department of Computer and Information Science, Temple University, Philadelphia PA, USA, dxj@ieee.org
3. School of Computer, Hangzhou Dianzi University, Hangzhou, China, perryliu@hdu.edu.cn

*Abstract*—**As the next generation of power system, smart grid provides people with great convenience in efficiency and quality. It supports the two-way communication and extremely improves the efficiency of utilization of energy resource. In order to dispatch accurately and support the dynamic price, a lot of smart meters are installed at user's house to collect the real-time data. However, all these collected data are related to user privacy. In this paper, we propose a data aggregation scheme based on secret sharing with error tolerance in smart grid, which ensures that the control center gets the integrated data without revealing users' privacy. Meanwhile, we also consider the differential privacy and error tolerance during the data aggregation. At last, we analyze the security of our scheme and carry out experiments to validate the results.**

## I. INTRODUCTION

As a new generation of energy network, smart grid is considered a useful way to solve the severe environment and resource problems. It is the combination of energy network and information technology. Differing from the traditional grid, the control mode of the smart grid is more flexible and reliable. It supports bidirectional power flow between the users and grid. To realize the optimal scheduling, smart grid installs a smart meter at each house to collect the real-time data .Not only that, smart grid also adopts many new service modes. For example, the dispatch center can make the dynamic price to encourage users to adjust their power consumption behaviors, collect the electricity consumption requirements and create the electricity generation plan in advance.

Although smart grid has many advantages in those aspects, there also exist several risks which may disclose user privacy. An adversary can infer user's family behaviors through his real-time data. For example, when you get up, when you go to work, when you take a shower and so on. Thus, thieves may gain entry to user's house when they notice that there is nobody home. Therefore, the privacy-preserving in smart grid becomes an extremely important problem which can hamper the implement of smart grid.

For the privacy-preserving in smart grid, many scholars have proposed various solutions. As we know, the privacy-preserving strategy can be divided into two ways. One is to hide the user's identity, the other way is to protect the user's real-time power data. In this paper, we adopt the Paillier cryptosystem which is a homomorphic encryption scheme to realize the data aggregation and use the secret sharing scheme to realize the differential privacy and realize the error tolerance through the substitution strategy. The process of our scheme is shown in figture1.

The rest of this paper is organized as follows. Section II introduces the related work. Section III shows the system model and design goals. In section IV, some preliminaries are given. In section V, our scheme is stated. In section VI, security analysis is given. In Section VII, the performance of our scheme is evaluated. In Section VIII, the paper is concluded.

## II. RELATED WORK

To protect the privacy of users in smart grid, many scholars have proposed various strategies. These strategies can be classified into two aspects: 1) protect user's privacy by masking the real identity; 2) protect user's privacy by masking their real-time data.

Some works are focused on masking user's identity. A simple solution adopting a trusted- party to manage the identity list is proposed in [1]. However, finding a trusted-party is not easy. Cheung proposes a scheme based on blind signature to solve the privacy-preserving and validity-authentication in [2]. It ensures that the verifier can authenticate sender's signature while has no information about his privacy. The downside of this scheme is that users should send their electricity data to the third party for authentication before communicating with the control center, which is against with the real-time property. Camenisch and Lysyanskaya propose a scheme named CL - Signature scheme which is similar to the blind signature in [3]. An effective scheme based on virtual ring is presented in [4]. It groups the users by their geographical positions and distributes each member in the same group with the same *ID*. In this way, control center can obtain all of the users' data without knowing the senders' *ID*. Obviously, it's a good way to protect user privacy, but the validity-authentication can't be guaranteed because of the anonymity. Privacy-preserving scheme based on pseudonym is also very common such as [5], and it always combines with the ring signature or blind signature to mask user's identity.

Some works are focused on masking user's real-time data. A solution using a battery to hide the real-time data is proposed in [6]. In this scheme, smart grid and the household battery provide users with electricity at the same time to mask the electricity data. The downside of this method is that the battery charging and discharging frequently may reduce the lifetime of battery. Data aggregation is also popular in smart grid for privacy-preserving. The Paillier encryption and Bone-Goh-Nission encryption are classical homomorphic encryption algorithms for data aggregation and they are used in many schemes such as [7]-[10]. However, the computational complexity of them can't be ignored. In [11], Jo tries to group the members to realize a distributed authentication in order to reduce the complexity, but the effect is not very ideal. The bilinear mapping is also a common solution for data aggregation. We usually use the bilinear mapping to create homomorphic encryption such as [12]. Remarkably, it is often used to realize the key-exchange. Secret sharing scheme is proposed to realize the data aggregation. It adopts the Shamir technique to encrypt the electricity data [13]. It divides a secret into different pieces and distributes them to various entities in smart grid. Only the control center acquires fixed number of shares, can he get the integral secret. Dong, Xiaolei, J. Zhou, and Z. Cao propose a simple solution to aggregate the real-time data in [14], it constructs an equation 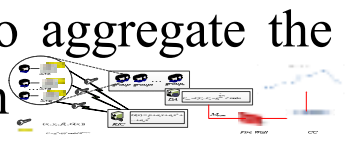 . All the *SM*s share the same value of $S$ , but it would increase the traffic in the network. In [15], Beussink, A shows a scheme based on data obfuscation, which adds a random number to each electricity data, but it will cause some large errors if the random numbers are not reasonable. In [7], Shi, Z and Sun, R discuss about the error tolerant and differential privacy. It groups all the users in smart grid and drops a group if this group has a malfunctioning smart meter. However, the error rate and computational complexity is not very ideal and we will compare our solution with this scheme at below.

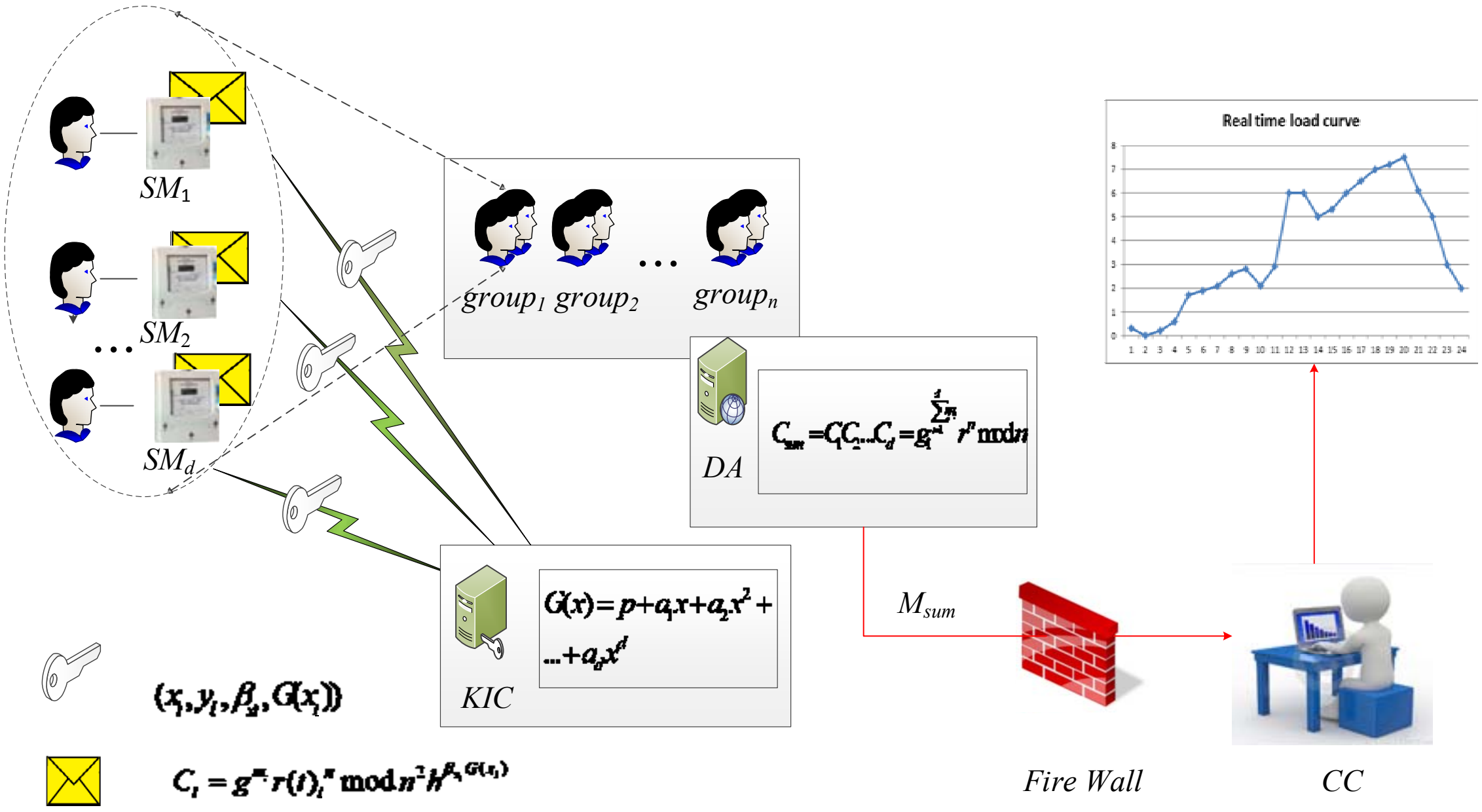

Fig.1 System Model

## III. SYSTEM MODEL AND DESIGN GOALS

### A. System model

As shown in figure 1, smart grid is divided into four parts, which are comprised of the control center (*CC*), the Key initialization center (*KIC*), the data aggregation device (*DA*), and residential users.

#### 1) Residential users

We divide all of the users into different groups in accordance with their geographical locations and the number of users in each group is the same. Each residential user is installed with a smart meter (*SM*) to collect the real-time data of the house applications .it can encrypt the data by paillier encryption and send these encrypted data to the data aggregation device.

#### 2) Data aggregation device

The data aggregation device is responsible for collecting all the data sent by *SM*s, calculating the sum of real-time data and uploading to the control center. In addition, it has a fault-tolerant function. When a smart meter is malfunctioning in a group, his data would be replaced by the other group's *SM* which has the same secret key.

#### 3) Key initialization center

The key initialization center is responsible for initializing all of the keys for *SM*s. It produces private keys by secret sharing scheme. Additionally, each group has the same number of smart meters, and the keys assigned to each group correspond to equal.

#### 4) Control center

The control center can acquire a summary of real–time data in smart grid. With these data, the *CC* can get the trend of power consumption and make the power generation plan or dynamic price immediately.

### B. Design goals

Considering the above scenarios, our design goals are as follows:

#### 1) Privacy-preserving.

A residential user's data are inaccessible to any other users. No matter the outside adversary or *CC* should not acquire the real-time data of users even if he knows the cipher text and encryption algorithms.

*2) Differential privacy*

Differential privacy refers to the problem of preventing users from compromising privacy during the data aggregation process. The common construction method is as follows.

For two formulas during the data aggregation:

$$C_{sum1} = C_1 C_2 \ldots C_n \tag{1}$$

$$C_{sum2} = C_1 C_2 \ldots C_n C_{n+1} \tag{2}$$

The data aggregation device can compute $M_{sum1}$ and $M_{sum2}$ by corresponding homomorphic encryption. Then, the plain text of the $SM_{n+1}$ will be disclosed easily through

$$M_{n+1} = M_{sum2} - M_{sum1} \tag{3}$$

We call this attack method differential attack, and the common solution is adding a random number into the data. However, this solution would increase some errors. Finding a novel solution which can protect differential privacy while has low error rate is our design goal.

*3) Error-tolerance*

Only the data aggregation device collects all the *SM*s' data, can he acquire the sum of the data. If there is a *SM* damaged, the data aggregation can't run in the right way. Therefore, when there is a malfunctioning *SM* in a group, we must ensure the data aggregation still run normally.

## IV. PRELIMINARIES

### *A. Notations*

Table1. Notations

| Acronym | Descriptions |
|---|---|
| *ID* | Identification number |
| SM | Smart meter |
| *CC* | Control center |
| *KIC* | Key initialization center |
| *DA* | Data aggregation device |
| $e_o$ | The error rate of our scheme |
| $e_D$ | The error rate of DG-APED |
| *t* | Time stamp |
| $M_i$ | Plaintext of $SM_i$ |
| $C_i$ | Ciphertext of $SM_i$ |
| *r* | Random number |
| *OS* | Our scheme |

### *B. Paillier cryptosystem*

Paillier cryptosystem is an asymmetric encryption algorithm, which has additive homomorphism properties. It includes three procedures: key generation, encryption and decryption.

*1) Key generation*

Chooses two prime numbers p, q and calculates $n = pq$ . $g$ is a generator of cyclic group $Z_{n^2}{}^*$, and $\gcd(L(g^{\lambda} \bmod n^2), n) = 1$ . The public key is $(n, g)$ , and the private key is $\lambda$ .

$$\lambda = lcm(p-1, q-1) . \tag{4}$$

*2) Encryption phase*

For the plain text $m \in Z_n$ , we can select a random number $r < n$ . Then, the ciphertext can be calculated as follows:

$$C = g^m r^n \bmod n^2 \tag{5}$$

*3) Decryption phase*

After receiving the cipher text *C*, the receiver can get the plain text *m* with the secret key $\lambda$ by the following formula

$$m = \frac{L(C^{\lambda} \bmod n^2)}{L(g^{\lambda} \bmod n^2)} \bmod n . \tag{6}$$

### *C. Secret sharing scheme*

The secret sharing scheme is a scheme which splits a secret into n pieces and distributes these pieces with different valid members. If an adversary captures a member in the system, he can only get a piece of the secret. Only if the adversary get at least k pieces of the secret, can he get the whole secret. We usually adopt a Shamir technique to realize this result.

The trusted-party chooses a polynomial to split a secret .

$$G(x) = p + a_1 x + a_2 x^2 + \ldots a_{d-1} x^{d-1} \tag{7}$$

$(x_i, y_i)$ is the corresponding share. Remarkably, the Shamir secret sharing scheme is the fully homomorphic and can be designed as a better scheme to realize the data aggregation. According to the lagrange interpolation polynomial, we have

$$G(x) = \sum_{j=0}^{d-1} \Big( \prod_{i=0, i \neq j}^{d-1} \frac{x_i - x}{x_i - x_j} \Big) G(x_j) \tag{8}$$

$$\beta_{x_i} = \prod_{j \neq i}^{d-1} \frac{x_j}{x_j - x_i} \tag{9}$$

Then, we can easily compute *p* as follows:

$$\sum_{i=0}^{d-1} G(x_i) \beta_{x_i} = G(0) = p \tag{10}$$

### *D. Log normal distribution*

Log normal distribution is a distribution in the mathematics field that many phenomena follow. Research [16] shows that the real-time data follow the log normal distribution. We can easily get the expectation by the probability density function of the distribution as follows:

The probability density function

$$p(x) = \frac{1}{x\sigma\sqrt{2\pi}} e^{-\frac{(\ln x - \mu)^2}{2\sigma^2}} \tag{11}$$

The expectation

$$E(x) = e^{\frac{\mu+\sigma^2}{2}} \quad (12)$$

## V. Our Scheme

### A. System initialization

The *KIC* first chooses two big primes $p,q$, and computes $n = pq$. Then, it chooses a generator $g_1$ from $Z_{n^2}^*$, where $\gcd(L(g_1^{\lambda} \bmod n^2), n) = 1$ and creates $h = g_2^q$, where $g_2 \in Z_n^*$ .After, it constructs a formula $G(x) = p + a_1 x + a_2 x^2 + ... a_{d-1} x^{d-1}$. All the *SM*s in smart grid are divided into different groups, and each group has the *d* members. For each *SM* in a group, the *KIC* assigns the private key $\{x_i, y_i, G(x_i), \beta_{x_i}\}$ to the $SM_i$, calculates $\lambda = lcm(p-1, q-1)$ and sends $\lambda$ to the DA. At last, it publishes $(n, h, H_1, H_2, g_1, g_2)$. While, $x_i$ is a random number representing the *SM* serial number. $y_i$ is the group serial number. $H_1$ and $H_2$ are two hash functions. Particularly, the set of *SM* serial numbers $\{x_1, x_2, ... x_d\}$ in each group is the same. That is to say, for a special *SM*, it can find *n*-1 members with the same private key except for the group serial number in other groups. When a *SM* receives the private key, it saves $\{x_i, y_i, G(x_i)\beta_{x_i}\}$ in his database.

### B. Encryption

The *SM* collects the electricity data every 15 minutes from all the house applications. For the time *t,* it computes

$$C_i = g^{m_i} r(t)_i^n \bmod n^2 h^{\beta_{x_i} G(x_i)} \quad (13)$$

$$H_1(t \mid G(x_i)\beta_{x_i}) \quad (14)$$

$$H_2(t \mid C_i \mid h(G(x_i)\beta_{x_i}) \quad (15)$$

Then, the *SM* sends $\{y_i, C_i, H_1(t \mid G(x_i)\beta_{x_i}), H_2(t \mid C_i \mid h(G(x_i)\beta_{x_i}))\}$ to the *DA*. It is worth noting that the serial number of *SM* $x_i$ should be kept secret.

### C. Data aggregation

When the *DA* receives a message from a *SM*, it verifies $H_2(t \mid C_i \mid h(G(x_i)\beta_{x_i})$ firstly. If the hash value is right, the data aggregation will be performed. Otherwise, the message will be dropped.

*1) Normal aggregation*

If the *DA* receives all the *SM*s' data in a group, it runs the data aggregation as follows:

$$C_{sum} = \prod_{i=1}^{d} C_i = g_1^{\sum_{i=1}^{d} m_i} (\prod_{i=1}^{d} r_i)^n \bmod n^2 h^{\sum_i^d \beta_{xi} G(xi)} = g_1^{\sum_{i=1}^{d} m_i} (\prod_{i=1}^{d} r_i)^n \bmod n^2 \quad (16)$$

$$M_{sum} = \frac{L(C_{sum}{}^{\lambda} \bmod n^2)}{L(g_1^{\lambda} \bmod n^2)} \bmod n \quad (17)$$

*2) Error tolerance*

If there is a malfunctioning *SM* in a group, the *DA* runs the following steps:

First, it compares this group of hash table constituted by $H_1(t \mid G(x_i)\beta_{x_i})$ with other complete groups to find the malfunctioning *SM.*

Then, selects a $SM_j$ from other groups with the same hash value $H_1(t \mid G(x_i)\beta_{x_i})$ as the malfunctioning $SM_i$ to replace. To further reduce the error, the data of $SM_j$ is processed before the data aggregation as follows:

$$\bar{m} = \frac{\sum_i^n M_i^{'}}{n} \quad (18)$$

$$c_j^{'} = \frac{c_j}{g_1^{\bar{m}}} = g_1^{m_j - \bar{m}} r^n{}_j \bmod n^2 h^{\beta_{xj} G(xj)} \quad (19)$$

$\sum_i^n M_i^{'}$ represents the sum of the electricity data of the previous period. $c_j^{'}$ represents the processed data of $c_j$ and replaces the missing data $c_i$ to run the data aggregation.

### D. Power dispatching

After receiving sum of the electricity data in smart grid, the *CC* can draw the real-time load curve and create the dynamic price, power generation plan and other scheduling strategies.

## VI. Security Analysis

### A Privacy-preserving

For a *SM* in smart grid, we can analyze the security of data from three aspects: external attacker, *DA*, *CC* and malicious user with the same private key.

*1) External attacker*

When an external attacker compromises the user's *SM*, the cipher text $C_i$ sent by the user can be obtained. However, because the attacker doesn't know the other d-1 users 'private key, it can't acquire the plain text due to the lack of *p*.

*2) DA*

After receiving all the data from *SM*s, the *DA* can only perform the data aggregation and replace the malfunctioning one when it's necessary. However, it can't obtain the decryption key $p$, therefore, the security of the user's data can be ensured.

*3) CC*

The *CC* can only get all the users' aggregated data, so it can't snoop to a single user's real-time data. Thus, user's privacy can be ensured in our scheme.

*4) Malicious user*

After receiving the cipher text *C*, the receiver can get the plain text *m* with the secret key $\lambda$ by the following formula:

$$m = \frac{L(C^{\lambda} \bmod n^2)}{L(g^{\lambda} \bmod n^2)} \bmod n \quad (20)$$

For a specific *SM* in a certain group, there are $n-1$ *SM*s in the remaining groups according to the design idea of our scheme. If a malicious user exists in this set, the privacy of the boon *SM* may be threatened. However, there is no security risk in our scheme. For two users with the same key in different groups,

$$C_i = g_1^{m_i} r(t)_i^{\,n} \bmod n^2 h^{\beta_x G(x)} \quad (21)$$

$$C_j = g_1^{m_j} r(t)_j^{\,n} \bmod n^2 h^{\beta_x G(x)} \quad (22)$$

Even if the malicious user knows the key $h^{\beta_{x_i} G(x_i)}$, as they do not know the paillier encryption algorithm decryption key $\lambda$ , so it still can't acquire the plaintext.

### B. Differential privacy

Differential attack needs to calculate the integrated data of $d+1$ users and the integrated data of $d$ users. After computing their difference, the attacker can acquire the privacy information of the individual user. For this scheme, there is no risk of differential attack. Because the secret sharing scheme can only calculate the *d* users' integrated data. If the number of *SM*s participating in the data aggregation is less than *d* or more than *d*, the key $p$ can't be synthesized.

### C. Error-tolerance

Data aggregation often involves fault tolerance. In this paper, when a certain *SM* is damaged, it is difficult to synthesize the key. Therefore, we adopt the idea of different group substitution to realize error tolerance. Because real-time data is collected for scheduling, creating the power generation strategies, and dynamic price, error caused by substitution is negligible for the big data in smart grid. When there are too many malfunctioning *SM*s in smart grid, the *KIC* should re-assign the keys.

## VII. PERFORMANCE EVALUATION

The most important idea of our paper is to use substitution to realize the error-tolerance, so it is necessary to prove the substitution between two members in different groups to be right. Because the real-time data follow the log normal distribution, we can use its expectation to calculate the error rate.

$$e_O = \frac{S-S'}{S} = \frac{MC_j{}'}{N\overline{C}} \approx \frac{M}{Ng^m} \text{(g=13)} \quad (23)$$

As smart grid is very huge, we suppose the number of *SM*s is from 4000 to 10000 and the number of malfunctioning *SM*s is from 0 to 200. By inputting the number of malfunctioning *SM*s and the total number of the *SM*s in smart grid into the formula (23), we can get the error rates 'curve as figure.2.

For the number of *SM*s is from 4000 to 10000, we can find that the error rate of our scheme is less than 0.07% in the worst case and the average rate is close to 0.02%, which is allowed in smart grid for the power scheduling. As the formula (23) shows, the error rate is related to the value of g .Therefore, we can increase the accuracy by adjusting the value of g .

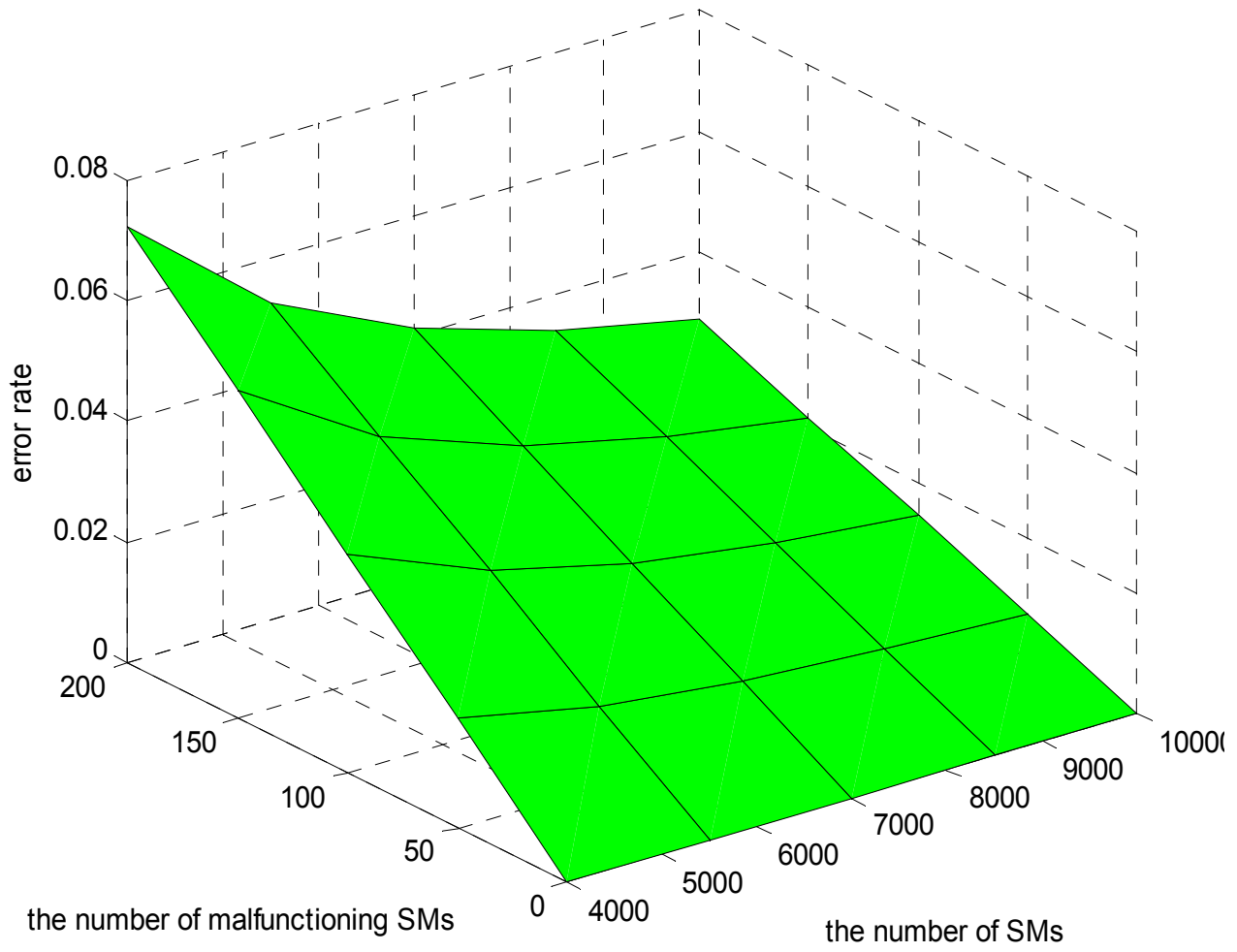

Fig.2 The error rate in our scheme

As far as we know, DG-APED scheme proposed by Shi, Zhiguo is also a subtle group-based scheme with error tolerance [7]. It adopts the data drop strategy to realize the error tolerance during the data aggregation. We calculated the computational complexity of the two schemes in error processing and show them in the figure.3.

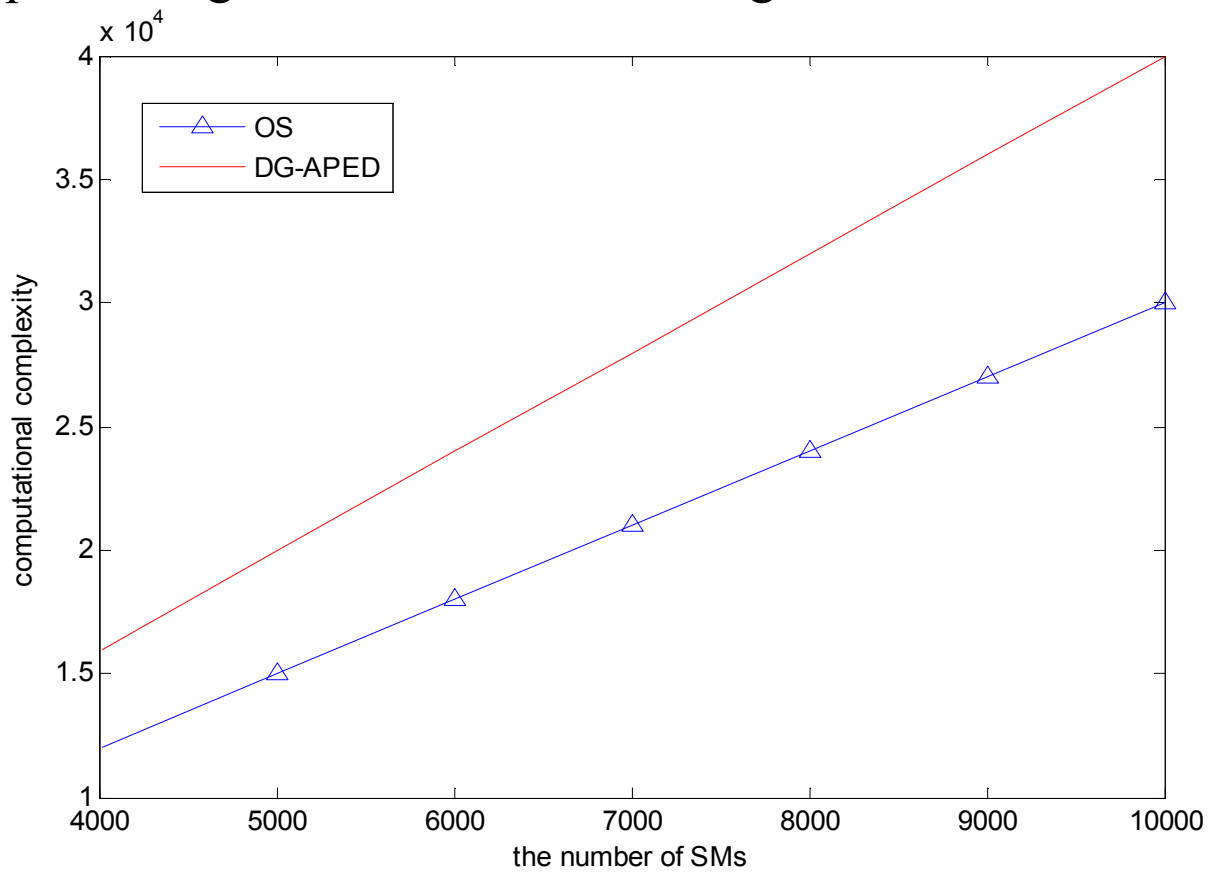

Fig.3 The computational complexity

In addition, we also calculated the error rate of the two schemes and show them in figure 4. Our scheme's error rate can be calculated as follows:

$$e_O = \frac{S-S'}{S} = \frac{MC_j{}'}{N\overline{C}} \approx \frac{M}{Ng^m} \text{(g=13)} \quad (23)$$

While, the error rate of the DG-APED can be calculated as follows:

$$e_D = 1-(1-\frac{M}{N})^d - (\frac{M}{N})^d \, (d=10) \quad (24)$$

Through the figure 4, we can find that when the malfunction rate (M/N) is less than 0.97, the error rate of our scheme would be much lower than DG-APED. While, when the malfunction rate is more than 0.97, the system has to reassign the keys for all working *SM*s. Therefore, our scheme has great advantage in the accuracy.

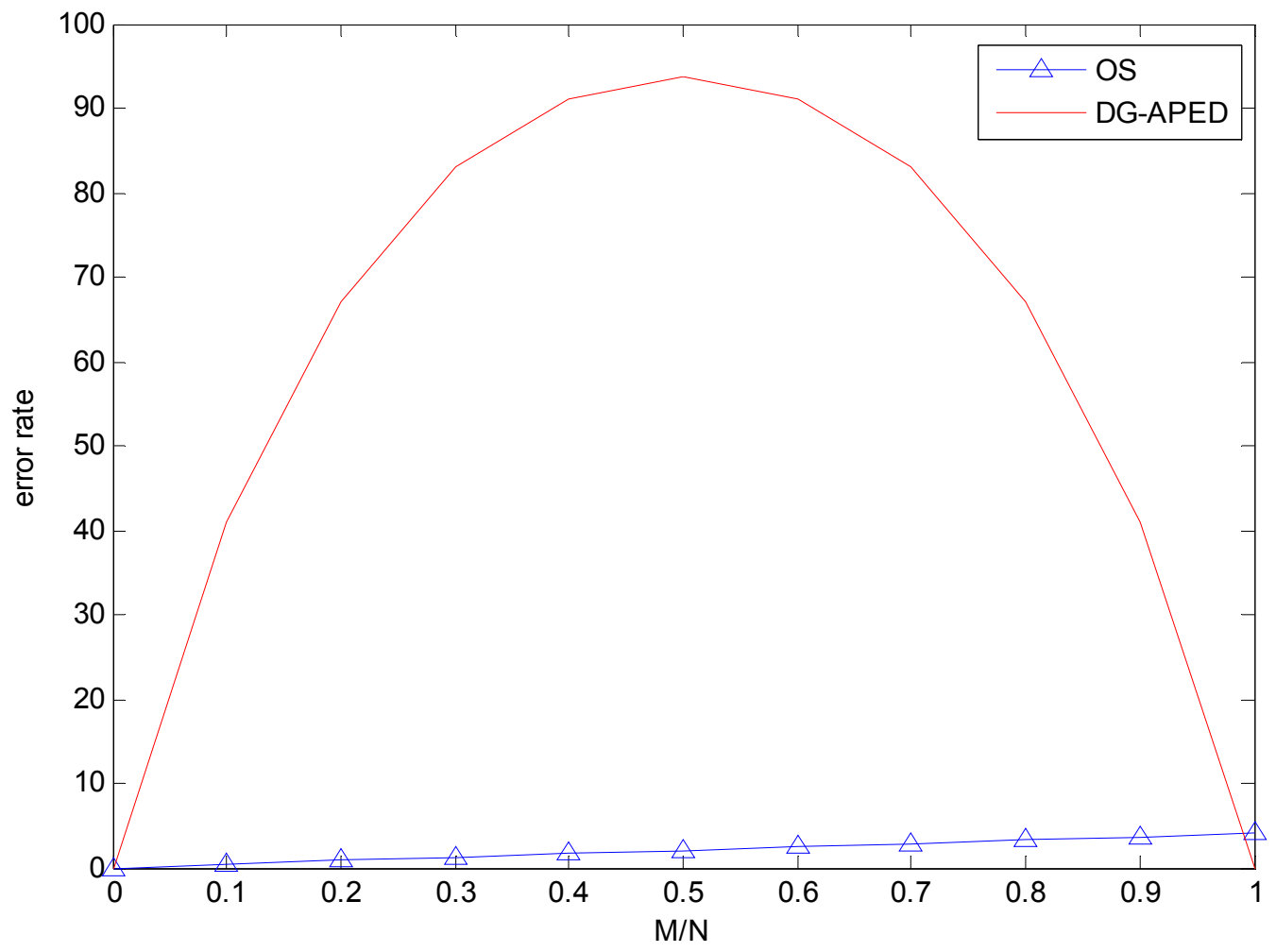

Fig.4 The error rate of two schemes

## VIII. CONCLUSION

Smart grid is an effective way to resolve the severe environment and resource problem. Unfortunately, it also causes some privacy-disclosing problems. In this paper, we propose a data aggregation scheme based on secret sharing scheme and paillier encryption. We achieve the error-tolerance for the normal aggregation by substitution. Therefore, even if there are some malfunctioning *SM*s in smart grid, our scheme can also run in a normal way. In the future work, we will focus on combining the real-time data privacy with the normal billing.

## ACKNOWLEDGMENT

This work is partially supported by Natural Science Foundation of China under grant 61402171, the Fundamental Research Funds for the Central Universities under grant 2016MS29, as well as by the US National Science Foundation under grant CNS-1564128.